\documentstyle[prb,aps]{revtex}
\tighten
\begin{document}
\preprint{Bicocca/FT-00-09  June 2000}
\draft

\title
 { Extension to  order $\beta^{23}$ of the high-temperature expansions\\
 for the spin-1/2 Ising model on the simple-cubic  and\\
 the body-centered-cubic lattices\\}
\author{P. Butera\cite{pb} and M. Comi\cite{mc}}
\address
{Istituto Nazionale di Fisica Nucleare\\
Dipartimento di Fisica, Universit\`a di Milano\\
Via Celoria 16, 20133 Milano, Italy}
\date{\today}
\maketitle
\begin{abstract}
Abstract: Using a  renormalized linked-cluster-expansion method, 
we have extended to  order $\beta^{23}$ the high-temperature series
for the   susceptibility $\chi$ and  
the second-moment correlation length $\xi$  of the 
 spin-1/2 Ising models
on the sc and the bcc lattices.
A study of these expansions 
 yields  updated direct estimates of  universal  parameters, such as  
 exponents and amplitude ratios, which characterize
 the critical behavior of  $\chi$ and $\xi$.
Our best estimates for the inverse critical temperatures are
 $\beta^{sc}_c=0.221654(1)$ and $\beta^{bcc}_c=0.1573725(6)$. 
For the susceptibility exponent
 we get $\gamma=1.2375(6)$ and for the correlation length 
exponent $\nu=0.6302(4)$.
 The ratio of the critical amplitudes of  $\chi$ above and below the 
critical temperature is estimated to be $C_+/C_-=4.762(8)$. The analogous 
ratio for $\xi$   is estimated to be $f_+/f_-=1.963(8)$.
For the correction-to-scaling amplitude ratio we obtain 
$a^+_{\xi}/a^+_{\chi}=0.87(6)$.  
\end{abstract}

%\narrowtext
 \widetext

\section{Introduction}
 As a part of an ongoing long-term  program of computer-based
 calculations and analyses of HT series for two-dimensional\cite{bc2d} and
 for three-dimensional\cite{bc3d,bcsd} lattice spin models,
we have  extended by two terms the high-temperature (HT) series for the
spin-1/2 Ising model  on the  simple-cubic (sc) and  the 
body-centered-cubic (bcc) lattices.
In the  first analysis presented here,
 we shall restrict to consider the HT expansions 
through $\beta^{23}$ for the susceptibility $\chi$  and the 
second-moment correlation length $\xi$, mainly in order to update the 
  {\it direct} 
estimates of the corresponding critical indices $\gamma$ and $\nu$.

For the sc lattice, the longest expansions of these quantities 
already in the literature 
 reach order $\beta^{21}$. They were obtained and analyzed
 in Refs.\cite{bc3d,bcsd} only a few years ago.
In the case of the bcc lattice, the published series\cite{nick21,nr90}
  for $\chi$ and $\xi$,  also extending through $\beta^{21}$, 
were calculated by B.G.Nickel two decades ago.
The  progress in such computations has been slow
 due to the exponential growth  
 of their complexity  with the order of the expansion, 
so that even adding only a few 
terms to the present results is a laborious task. 
 Within the   renormalized 
linked-cluster\cite{lk} expansion method,
 used in our work, one must  overcome many 
problems of combinatorial nature concerning 
graph generation, classification and 
partial resummation,  
 and a special effort must be devoted to keep
 under strict control the numerous possible sources of error.	
 In our case, a final severe test is provided by having the program 
to reproduce, in three dimensions,  established data like 
the series for the nearest-neighbor spin 
correlation on the sc lattice, which is already 
tabulated\cite{cor25} through $\beta^{27}$,
and, in two dimensions,  the series for $\chi$  and  
 $\mu_2$ on the simple square lattice, which are known\cite{nick21,nr90}  
through $\beta^{35}$  and beyond\cite{guttse,garten}.
After the completion of this work, a preprint\cite{campos}
 has been issued which also
reports independently extended 
expansions for $\chi$ and $\mu_2$ on both the 
the sc and the bcc lattice through orders  $\beta^{23}$ and  $\beta^{25}$
 respectively. Our series coefficients 
agree with those of Ref.\ \cite{campos} 
as far as the expansions  overlap. This adds further confidence about the 
correctness of the results since our implementation of the linked cluster 
expansion procedure is rather different 
from that described in Ref.\cite{campos}.

Any enrichment of the exact informations on the 3d 
 Ising model is still of general 
interest. Here  we have used these novel data  to improve the 
knowledge of the nonnegligible singular
 corrections\cite{wegn} to the leading critical 
singularities of $\chi$  and  $\xi$  and, as a consequence, the 
accuracy of the direct HT series estimates of all critical parameters.
As stressed in Ref.\cite{nick21,nr90}, 
the corrections to scaling first showed up unambiguously when
 the bcc series were extended to order  $\beta^{21}$, the 
last three coefficients being crucial\cite{nr90}. 
It is therefore 
 helpful to produce more coefficients, in order to stabilize and possibly
 refine the quality of the information extracted from the series.

The plan of this note is as follows:
after setting our notational conventions in Sec. II, we tabulate the series
 coefficients for $\chi$  and  $\mu_2$ through  order $23$, with 
respect to  the usual HT expansion  variable   $v=th(\beta)$.
In Sec. III we  report the results of our extrapolations 
for the critical temperatures,  
for the critical 
exponents $\gamma$ and $\nu$, for the universal ratio $C_+/C_-$  of the 
 critical amplitudes of the susceptibility 
above and below the critical point, 
 for the analogous ratio $f_+/f_-$ of the correlation-length amplitudes 
and for the ratio $a^+_{\xi}/a^+_{\chi}$ of 
the correction-to-scaling amplitudes\cite{aha}. 
  Our estimates are  compared with the latest numerical 
calculations by series,  by stochastic methods and by perturbative 
renormalization group (RG) techniques,  in the fixed-dimension (FD) 
 approach\cite{zib,mur,ant,kle99,guida,zinn,itzy} and in 
the $\epsilon$-expansion approach\cite{guida,zinn,itzy,kleb,epsi,epv}.
Less recent studies have been already 
reviewed in our Refs.\cite{bc3d,bcsd}.

\section{ Definitions and notations}
In order to introduce our notation,  we shall specify by 
the Hamiltonian

\begin{equation}
H \{ s \} = -{\frac {J} {2}} \sum_{\langle \vec x,{\vec x}' \rangle } 
s({\vec x})  s({\vec x}')
\label{hamilt} \end{equation}
 
the nearest-neighbor three-dimensional spin-1/2 Ising model in zero magnetic 
field.

Here $s(\vec x)=\pm 1$ is the  spin variable at the
lattice site $\vec x$,   
and the sum extends over all nearest neighbor pairs of  sites. 
We shall consider expansions in the usual HT variable $\beta=J/k_BT$
 called  ``inverse temperature'' for brevity.
 However, for convenience, we shall tabulate the series coefficients with 
respect  to the expansion variable  $v=th(\beta)$.

The susceptibility is expressed in terms of 
the connected two-spin correlation
 function  $\langle s(\vec x) s(\vec y) \rangle_c$  by

\begin{equation}
\chi(\beta) = \sum_{\vec x} \langle s(0)  s(\vec x) \rangle_c = 
 1+ \sum_{r=1}^\infty a_r \beta^r ; \label{chi} \end{equation}

and the second  moment of the correlation function is defined as

\begin{equation}
 \mu_{2}(\beta)=\sum_{\vec x} \vec x^2 \langle s(0)  s(\vec x) 
\rangle_c = 
 \sum_{r=1}^\infty b_r \beta^r. 
\end{equation}

In terms of $\chi$ and $\mu_{2}$  the 
 second-moment correlation length $\xi$ is  defined by 

\begin{equation}
 \xi^{2}(\beta)= \frac  {\mu_{2}(\beta)} {6\chi(\beta) }. 
\end{equation}

For easy reference we report here the complete expansions 
 of  $\chi$ and $\mu_{2}$,  rather than only the 
lastly computed two coefficients. For the susceptibility on the 
sc lattice we have 

\scriptsize
\[ \chi^{sc}(v)=1 + 6v^{}+ 30v^{2}+150v^{3}+726v^{4} +
 3510v^{5}+ 16710v^{6}+79494v^{7}+ 375174v^{8} +   
1769686v^{9}+ 8306862v^{10}\]\[+38975286v^{11}+182265822v^{12} +
852063558v^{13}+ 3973784886v^{14}+ 18527532310v^{15}  + 
86228667894v^{16}\]\[+ 401225368086v^{17}+ 1864308847838v^{18} + 
8660961643254v^{19}+40190947325670v^{20}+ 186475398518726v^{21} \]
\[+ 864404776466406v^{22} + 4006394107568934v^{23}+ \ldots \]

\normalsize
for the second moment on the sc lattice:   
\scriptsize

\[\mu_2^{sc}(v)= 6v^{}+72v^{2}+582v^{3}+4032v^{4}+25542v^{5}+153000v^{6}+
880422v^{7}+ 4920576v^{8}+ 26879670v^{9}+ 144230088v^{10}\]\[+   
762587910v^{11}+
3983525952v^{12}+20595680694v^{13}+105558845736v^{14}+536926539990v^{15}+
2713148048256v^{16}\]\[+ 13630071574614v^{17}+ 68121779384520v^{18}+
338895833104998v^{19}+ 1678998083744448v^{20}+ 8287136476787862v^{21} \]\[+
 40764741656730408v^{22} + 199901334823355526v^{23} + \ldots  \]

\normalsize

for the susceptibility on the bcc lattice: 
\scriptsize 
\[ \chi^{bcc}(v)=1 + 8v^{}+56v^{2}+  392v^{3}+ 2648v^{4}+ 17864v^{5}+ 
118760v^{6}+ 789032v^{7}+ 5201048v^{8}+ 34268104v^{9}+224679864v^{10}\]\[+
1472595144v^{11}+ 9619740648v^{12}+  62823141192v^{13}+ 409297617672v^{14}+
2665987056200v^{15}+ 17333875251192v^{16}\]\[+ 112680746646856v^{17}+
731466943653464v^{18}+ 4747546469665832v^{19}+ 30779106675700312v^{20}+
199518218638233896v^{21} \]\[ 
+ 1292141318087690824v^{22}+ 8367300424426139624v^{23}
 +\ldots    \]

\normalsize
for the second moment on the bcc lattice:  
\scriptsize 

\[\mu_2^{bcc}(v)=  8v^{} +128v^{2}+1416v^{3}+13568v^{4}+ 119240v^{5}+
992768v^{6}+ 7948840v^{7}+61865216v^{8}
+470875848v^{9}+ 3521954816v^{10}\]\[+
25965652936v^{11}+ 189180221184v^{12}
+ 1364489291848v^{13}+9757802417152v^{14}+69262083278152v^{15}
+488463065172736v^{16}\]\[+3425131086090312v^{17}+
23896020585393152v^{18}+165958239005454632v^{19}
+1147904794262960384v^{20}\]\[+
7910579661767454248v^{21}+ 54332551216709931904v^{22} 
+ 372033905161237212392 v^{23} + \ldots \]

\normalsize 

\section{ Analysis of the series} 

In terms of the reduced inverse 
temperature $\tau^{\#}=1- \beta/\beta^{\#}_c$,
 the asymptotic critical behavior of the 
susceptibility is expected to be\cite{wegn} 
\begin{equation}
 \chi^{\#}(\beta)
\simeq C_+^{\#}(\tau^{\#})^{-\gamma}\Big(1+ a^{+\#}_{\chi}
(\tau^{\#})^{\theta}+ \ldots +e^{+\#}_{\chi}\tau^{\#} +\ldots \Big)
\label{conf}\end{equation} 

as the critical point $\beta^{\#}_c$ is approached from below.
(Here and in what follows,  the superscript ${\#}$ stands 
for either sc or bcc, as appropriate, and 
 will be dropped whenever  unnecessary. The index +(-) denotes, as usual, 
 quantities associated with the high (low) temperature 
side of the critical point.)
Similarly, for the correlation length $\xi$,  we expect 	

\begin{equation}
 \xi^{\#}(\beta)
\simeq f_+^{\#}(\tau^{\#})^{-\nu}\Big(1+ a^{+\#}_{\xi}
(\tau^{\#})^{\theta}+ \ldots +e^{+\#}_{\xi}\tau^{\#}+\ldots \Big)
\label{confx}\end{equation} 
 as $\tau \rightarrow 0^+$.

The exponents $\gamma$, $\nu$ and $\theta$ are universal 
quantities, whereas 
the critical amplitudes $C_+^{\#}$,  $f_+^{\#}$,
the amplitudes $a^{+\#}_{\chi}$,  $a^{+\#}_{\xi}$ of the leading nonanalytic 
correction-to-scaling terms and the amplitudes $e^{+\#}_{\chi}$,  
$e^{+\#}_{\xi}$ of the leading analytic corrections
 are nonuniversal, as suggested by the superscript
${\#}$.
 Experimentally accessible  universal combinations can be formed 
out of the critical amplitudes\cite{aha}.
 Here we shall be concerned with  series estimates of
 the universal ratios $C_+/C_-$,  $f_+/f_-$ 
and  $a^{+}_{\xi}/a^{+}_{\chi}$. Notice that for the critical amplitudes 
we have adopted the notation of Ref.\cite{zinn} and of other recent studies 
 rather than that of Ref.\cite{aha}.

\subsection{Estimates of the critical points} 
As a first step of the analysis, we shall examine the series for 
the susceptibility whose coefficients have 
the smoothest pattern of behavior, so that they are 
 generally 
used to estimate the critical temperatures. These estimates will also be used 
to bias the determination of the critical exponents and of the universal 
amplitude ratios, therefore their 
accuracy is crucial.
 Let us begin by considering the results obtained by a very 
 efficient   variant of the ratio method introduced by 
J. Zinn-Justin\cite{zinn79}, (see also\cite{guttse}).

We evaluate $\beta_c$ from the sequence 
\begin{equation}
 (\beta_c)_n = (\frac {a_{n-2}a_{n-3}} {a_{n}a_{n-1}})^{1/4} 
exp[\frac { s_n+  s_{n-2}} {2 s_n( s_n- s_{n-2})}]=\beta_c 
+O(\frac{1} {n^{1+\theta}})
\label{betazinn}\end{equation} 
where 
\begin{equation}
 s_n=\Big( ln(\frac{a_{n-2}^2} {a_{n}a_{n-4}})^{-1} 
+ln(\frac{a_{n-3}^2} {a_{n-1}a_{n-5}})^{-1} \Big)/2
\end{equation} 
 This is an {\it unbiased} method, in the sense that 
 no additional 
accurate information must be used together with the series in 
 order to get the estimates of the critical parameters, but we found  
 useful to  improve the procedure by biasing it 
with the value of $\theta$ as follows.
 For sufficiently large $n$, the sequence of estimates $(\beta_c)_n$ 
shows  very small regular oscillations due to the 
loose structure of the lattice. 
Moreover the odd and even subsequecences of  $(\beta_c)_n$  have
 a residual decreasing trend which is very nearly 
linear on a $1/n^{1+\theta}$ plot, as suggested by 
eq. (\ref{betazinn}).
 Therefore, simply taking the highest order term of the 
sequence $(\beta_c)_n$
 as the final estimate, would be an inadequate choice. 
   We have preferred to extrapolate   separately  to 
$n \rightarrow \infty$ 
the  successive odd and  even 
 pairs of estimates
 $(\beta_c)_n$, 
 assuming that we know the value of $\theta$ well enough. 
The two sequences of extrapolated values need further extrapolation 
 which allows also for the 
 small residual curvature of the plot and leads to the final 
 estimates $\beta^{sc}_c=0.221654(1)$, in the case of the sc 
lattice,  
and $\beta^{bcc}_c=0.1573725(6)$, in the case of the bcc lattice.
The errors we have reported, account generously both 
for  the present uncertainty in $\theta$ (whose effects in  
this analysis are very small anyway)
 and for the uncertainty of the second  extrapolation.
 For the correction-to-scaling exponent we have assumed
   the value $ \theta=0.504(8)$,  
 obtained by the FD perturbative RG \cite{guida}. Also in the rest of 
 this note the central 
 values of all $\theta$-biased estimates  will refer to this  value.   
However, in the calculations of this and the next subsection, we  have also
considered a much larger uncertainty, 
 in order to make sure that our  results are compatible  
 with  somewhat higher central values 
 such as  $\theta=0.52(3)$, proposed in Ref.\cite{nr90} 
(as well as in  Ref.\cite{blh}, with a smaller error),  or 
 with  $\theta=0.53(1)$
 suggested in Ref.\cite{has}. An even larger central value
 $\theta=0.54(3)$ was 
 indicated in Refs.\cite{fisher,zinnfish}, while 
an experimental measure reported in Ref.\cite{henke} yields $\theta=0.57(9)$. 
 In the case of 
the bcc lattice, as an example of our extrapolation procedure, 
we have reported in Table 1 the 
last eight terms of the sequence $(\beta_c)_n$ and 
the results of the initial extrapolation
 of the last six successive alternate pairs of terms. 
 Our final result for the critical 
inverse temperature of the Ising model on the
 sc lattice is completely compatible, although 
much less precise than the value
 $\beta^{sc}_c =0.22165459(10)$ obtained from an extensive 
MonteCarlo (MC) study by a dedicated Cluster Processor\cite{blh} and
generally considered as the best  available estimate.  
 Our central value of  $\beta^{bcc}_c$, obtained similarly, is only
 slightly  smaller, but more precise 
than the value $\beta^{bcc}_c=0.157373(2)$  suggested in Nickel and Rehr 
analysis\cite{nr90,zinn79,george}. 
We should finally  mention that Prof. D. Stauffer\cite{stau} 
kindly informed us that
he still tends to favor the somewhat larger central estimate 
 $\beta^{sc}_c =0.221659$,  basing on the HT 
analysis in Ref.\cite{salm}, as well
 as on  his own recent simulation\cite{stau}
 of the critical dynamics and on analogous work in 
 Ref.\cite{ito}. We also recall that 
 a similar value  $\beta^{sc}_c =0.2216595(26)$ was indicated a decade ago in 
 the MonteCarlo simulation of 
Ref.\cite{land}. In the context of our analysis, these values
 lie approximately halfway between the highest order approximant 
$(\beta^{sc}_c)_{23} \approx 0.221667$ and 
our final estimate obtained from extrapolation. 
To close this section, three remarks are in order. First: 
the reliability of our analysis procedure 
 has been  corroborated by repeating it
 with the recently computed $O(\beta^{26})$ series 
for the self-avoiding-walk (saw) model on the sc lattice\cite{guttp}.
This is a relevant test because the structure of the corrections to scaling 
(namely the sign and size of the 
 correction amplitude and the value of the 
confluent exponent\cite{bcsd}) is expected to be quite
similar to the Ising sc case. For the saw model we have observed
 that the central value for $\beta_c$ indicated by our procedure is
essentially stabilized after reaching the order $\beta^{23}$ and 
 agrees closely with that indicated in 
 Ref.\cite{guttp}, while
 the error decreases as higher order 
coefficients are included in the analysis.
 Our procedure has also been tested and confirmed by 
other arguments in Ref.\cite{guttp}.
 Second: due to the higher coordination 
number of the bcc lattice, the corresponding  series have 
a greater  ``effective length''  than the sc series,
 and therefore all estimates obtained for the bcc
 lattice will be systematically more accurate. 
 Third: as expected, the inclusion in our 
analysis of the two additional coefficients for the expansion of 
$\chi$ on the bcc lattice, computed in Ref.\cite{campos}, does not 
essentially modify
 our central estimate of $\beta_c^{bcc}$, but only reduces its uncertainty
 to the value reported here. 

\subsection{Estimates of the critical exponents} 
By  using a related variant\cite{guttse,zinn79} of the ratio method 
 and by analogous arguments, 
 fairly good estimates can be 
obtained also for the exponents $\gamma$ and $\nu$. We construct 
 the approximation sequence
 \begin{equation}
 \gamma_n = 1 + \frac{2( s_n+  s_{n-2})} {( s_n- s_{n-2})^2}=
\gamma+O(\frac{1} {n^{\theta}})
\label{gammazinn}\end{equation} 
 with the same definition  as above for $ s_n$.
 Also in this case, for sufficiently large $n$, 
the successive estimates $\gamma_n$, (as well as the analogous 
ones $\nu_n$ obtained
from the series coefficients of $\xi^2$),  appear to be nearly linear on a 
 $1/n^{\theta}$ plot,  and therefore 
 we can follow an extrapolation procedure completely analogous to the one 
previously described. 
However, in the exponent calculation,
 the corrections are {\it a priori} larger 
 and therefore the procedure involves relative errors 
   larger than in the case of 
$\beta_c$. In order to illustrate this numerical procedure in the case of 
the bcc lattice, we have reported in Table 1 the 
last eight terms of the sequence $\gamma_n$ and 
the results of the extrapolation
 of the last six successive alternate pairs of terms. 
 The estimates  inferred from the analysis of these data are:
$\gamma=1.2378(10)$  and $\nu=0.629(2)$  in the case of the sc series and 
$\gamma=1.2373(6)$   $\nu=0.629(1)$ from the bcc series. 
 As expected, the relative uncertainties for the exponent $\nu$ are  
larger because of the slower 
approach of the second moment series to its asymptotic behavior.
The dependence of these estimates on the  value of $\theta$ used in the 
 extrapolation can be 
expressed as follows: $\gamma=1.2378+0.016(\theta-0.504)\pm 0.0010$ and 
$\nu=0.629 +0.02(\theta-0.504)\pm 0.0020$ in the case of the sc lattice;  
 $\gamma=1.2373+ 0.012(\theta-0.504) \pm 0.0006$ and $\nu=0.629 
+0.016(\theta-0.504)\pm 0.0010$ in the case of the bcc lattice.

In order to confirm these estimates for the exponents,
we shall resort also to (unbiased and biased) analyses by 
inhomogeneous differential approximants (DA's)\cite{guttse,da}.
  By unbiased DA's, we obtain somewhat larger  
estimates both for the critical inverse temperatures 
and for the exponents, which, however, show a clear decreasing trend.
Therefore also these data should be further extrapolated, 
but, unfortunately,  this is not as straightforward
 as in the case of the Zinn-Justin method.
 Thus we did not insist on this route and preferred to  
 perform   {\it biased} series analyses, either i) by 
the  first-order  {\it simplified}
 differential approximants (SDA) introduced 
and discussed in Ref.\cite{bcsd}, 
 in which both $\beta_c$ and  the correction-to-scaling 
exponent $\theta$ are fixed, 
 or alternatively ii)    by conventional second order inhomogeneous  DA's,  
 in which    $\theta$ and $\beta_c$ are varied in a small neighbourhood of 
their expected values, following the method of Ref.\cite{nr90}.
Let us also add that in all cases in which
 we have relied on SDA's, we have also
repeated the same calculation, 
first  subjecting the series to the biased variable change 
 introduced  by R. Roskies\cite{rosk} in order to regularize the leading 
correction to scaling and then computing simple Pad\'e approximants. 
 In this way we have always
obtained completely consistent results, although they are 
sometimes affected by larger uncertainties.

 We have used the procedure i) to study the residue of the log-derivative of 
$\chi$ or of $\xi^2$ at the critical singularity.
 In the case of the sc lattice series, 
 rather than our own estimate of  $\beta_c$, we have used
 the  more accurate (but otherwise completely consistent)
 value $\beta^{sc}_c =0.22165459(10)$ of  Ref.\cite{blh}.
  Thus we  estimate  $\gamma=1.2378(10)$ and  $\nu=0.6306(8)$.

In the analysis of the  bcc lattice series,  we have taken as a bias 
the value suggested by our extended ratio-method analysis
$\beta^{bcc}_c =0.1573725(6)$. 
 In this case we
 get the values $\gamma=1.2375(6)$ and $ \nu=0.6302(4)$.
   By using Fisher scaling law\cite{fisca}, 
we get $\eta=0.037(3)$ from the sc 
series and $\eta=0.036(2)$ from the bcc series.
 For both lattices we have used the same value 
(and uncertainty) of $\theta$ as 
 previously  discussed and  we have  easily allowed
 for the  residual decreasing trend of 
the exponent estimates, because SDA's values
 show a smaller spread than DA's.
We can also mention that in the bcc lattice case,  
the linearized dependence of the exponent 
 central estimates  on the bias values of $\beta_c$ and $\theta$ 
can be described as follows :
$\gamma=1.2375 +0.01(\theta-0.504) +90.(\beta_c-0.1573725)$ and 
$ \nu=0.6302  +0.015(\theta-0.504) +40.(\beta_c-0.1573725)$. 

 We shall take as our final estimates for the exponents 
those obtained by SDA's from 
the bcc lattice, which are best converged.

 The so-called M2 method of Ref.\cite{adler} is a very useful 
extension of the above mentioned Roskies' procedure\cite{rosk}. 
In the case of the bcc lattice it suggests 
$\beta^{bcc}_c =0.1573720(4)$ with $\gamma=1.2374(4)$ and $\theta=0.56(3)$, 
in good consistency with the other approaches.
On the other hand, in the case of the sc lattice,
 the results of the M2 method at 
order $\beta^{23}$, namely $\beta^{sc}_c =0.221659(2)$ $\gamma=1.2395(5)$
 with $\theta=0.50(2)$ are not essentially changed with respect to those 
obtained in Ref.\cite{salm} from the analysis of our previous 
$O(\beta^{21})$ series.  

In conclusion, provided that the sequences of estimates 
are carefully extrapolated using the independently computed value
 of $\theta$, the determination of the exponents by the 
 improved ratio method and by biased DA's or SDA's   are 
completely consistent, though the latter method gives 
 slightly more accurate results. 
At this order of expansion, asymptotic trends seem to be already 
stabilized and the uncertainties in the HT series estimates
 are   significantly reduced. A sample of  recent estimates of the critical 
exponents is reported in Table 2 and briefly commented in the rest of this 
subsection.
The agreement of our results 
 with the values $\gamma=1.2396(13)$ and $ \nu=0.6304(13)$, 
indicated by the FD perturbative RG\cite{guida}, 
or the values 
$\gamma=1.2380(50)$ and $ \nu=0.6305(25)$,
 suggested by the $\epsilon-$expansion\cite{guida}, is still
 good. However, we should observe that, in the years, 
 as the length of the HT series has increased,  the exponent
 estimates have been moving towards the slightly lower central values
 $\gamma \approx 1.237$ and $ \nu \approx 0.630$.
 Indeed very similar  values had already been suggested 
 some time ago by J.H. Chen, M. E. Fisher and B. G. Nickel\cite{fisher}  who 
 studied  soft spin models of the Ising universality class, chosen 
so to have negligible amplitudes for the leading corrections to scaling.
 Within this approach, the analysis\cite{nr90} of HT series through 
order $\beta^{21}$ for the bcc lattice gave 
$\gamma=1.237(2)$ and $ \nu=0.6300(15)$. A study\cite{campo} of the 
 HT series through $O(\beta^{20})$
for the sc lattice,  along the same lines as in Refs.\cite{nr90,fisher}, 
indicates $\gamma=1.2371(4)$ and  $ \nu=0.63002(23)$.  Recently, this method
 was adapted also to MC simulations in Ref.\cite{blh}, which reports 
$\gamma=1.2372(17)$ and $ \nu=0.6303(6)$. Analogously in Ref.\cite{has}, 
the estimates  $ \nu=0.6296(7)$ and $\eta= 0.0358(9)$ are  obtained, implying
$\gamma=1.2367(20)$.  Even lower
 central estimates of the exponents, 
namely $\gamma=1.2353(25)$ and  $ \nu=0.6294(10)$
 have  been obtained in a MC simulation of 
the Ising model by a finite-size 
scaling analysis\cite{balle} which allows for 
the corrections to scaling.              

\subsection{Estimates of  universal amplitude ratios} 

  By taking advantage
 also of the low-temperature expansion of $\chi$ on the sc lattice, 
extended in Ref.\cite{cor25} to order $u^{26}$, 
 (here $u=\exp(-4\beta)$),  and of the older series 
for the bcc lattice computed to order $u^{23}$  
in Ref.\cite{syk}, we can give a new direct  estimate  of the 
universal ratio $C_+/C_-$.  Using  the low-temperature 
series for $\xi^2$, computed for the sc lattice 
in  Ref.\cite{ari} through  
$u^{23}$, we can also compute  the ratio $f^{sc}_+/f^{sc}_-$.
 These quantities have been repeatedly evaluated in 
recent years by various techniques, with increasing accuracy.

 We have used  first order SDA's  to compute 
$C^{\#}_{\pm}=\lim_{\tau \to 0^{ \pm} }\vert \tau \vert^{\gamma}\chi^{\#}$.
 In the sc lattice case, by choosing $\beta^{sc}_c=0.22165459$, 
$ \gamma=1.2375$
 and $\theta=0.5$, we obtain $C^{sc}_+/C^{sc}_-= 4.762(8)$. 
Here the uncertainty refers to the 
sharp bias values 
above indicated. In order to compare this result 
 with others obtained by slightly different assumptions, the dependence of 
our estimate on the bias values $\gamma_b$ and $\theta_b$ for the critical
 and the correction exponents can be linearly approximated by 
$C^{sc}_+/C^{sc}_-= 4.762 +6.06(1.2375-\gamma_b)+ 0.7(\theta_b-0.5) \pm0.008$.
The ratio is  insensitive to the choice of $\beta^{sc}_c$ 
within its quoted uncertainty.
In the case of the bcc lattice, we have used 
the same bias values for $\gamma$ 
and $\theta$ together with
 $\beta^{bcc}_c=0.1573725$, obtaining $C^{bcc}_+/C^{bcc}_-= 
4.76(3)$.  In this case the error  (mainly coming from the 
uncertainty of $C^{bcc}_-$) is larger, 
but the result is completely consistent with the  sc 
lattice estimate. 

 The experimental measurements of this ratio
 range between 4.3 and 5.2\cite{aha,guida,zinn} and are 
perfectly compatible with our estimates. Other recent numerical 
evaluations are summarized in Table 3. However some comments are helpful for 
understanding these results. 
 The previous  evaluations by 
M.E. Fisher and coll. \cite{zinnfish,fietal}, 
 used  shorter series and 
 bias values $\gamma=1.2395$ and $\beta^{sc}_c=0.221630$ somewhat different
 from ours, thus 
yielding the slightly larger value 
4.95(15). 
The ratio  $C_+/C_-$ can also be obtained from  approximate parametric
 representations of the 
 scaling equation of state\cite{guida,campo,fisvar}: here we  quote only 
the most recent\cite{campo} such estimate:  4.77(2).

 The MC simulation of Ref.\cite{ruge} gave
the somewhat larger value 5.18(35), while the more recent and 
higher precision study of Ref.
\cite{case} yields 4.75(3) and the work of Ref.\cite{enge} reports 4.72(11).

 Within the $\epsilon-$expansion approach to the RG, the estimate 4.73(16) is 
obtained, while the FD expansion gives the result
 4.79(10)\cite{guida}. The value 4.72(17) was obtained in Ref.\cite{heit}.
 
In a similar way, we have computed 
 $f^{sc}_+/f^{sc}_-=1.963(8)$,  assuming 
$\nu=0.6302$.  The quoted uncertainty allows also for the 
uncertainties in the estimates of $\nu$ and $\theta$. 
 Other estimates appearing in the recent literature are summarized in 
Table 4.
 Our result compares 
well with the  estimate $1.96(1)$ obtained
in Ref. \cite{zinnfish,fietal} by shorter series
 as well as with the recent estimate  1.961(7) of 
Ref.\cite{campo}. The MC estimate of Ref.\cite{ruge} was 
$2.06(1)$, whereas in Ref.\cite{case} the value $1.95(2)$ is reported.
The latest $\epsilon-$expansion estimate\cite{zinn} 
is $1.91$ (with no indication of error bars)
 and the FD estimate\cite{heit} is $2.013(28)$. 
 The recent experimental estimates of this ratio range between 
 1.9(2) and 2.0(4). 

We believe that the close agreement between our series 
 estimates and the latest
determinations of these universal ratios is  due 
to the careful allowance   of the confluent 
corrections to scaling by SDA's. 
Indeed, even  using the longer series 
presently available, simple Pad\'e approximants,  notoriously inadequate to 
describe the singular 
corrections to scaling,  suggest estimates sizably larger, 
 while the conventional DA's lead to  a wider spread in the estimates.
 Further improvements of the direct series 
determination of these ratios should probably 
 await for an extension of the low-temperature series.

 By using only the HT extended series presented here, we can also reevaluate
 the universal ratio $a^+_{\xi}/a^+_{\chi}$. Let us 
recall that, as observed in Ref.\cite{bcsd,nr90,fisher,george} 
and  argued
 in earlier studies\cite{liufi}  
 for the spin-1/2 Ising model on the sc, the bcc and the fcc lattices, 
the  amplitudes of the leading  correction-to-scaling 
terms have a negative sign,  both for the susceptibility and the 
correlation length.  
The values of these amplitudes can  be  most simply  determined,
 also in this case, by using the
 SDA's  above mentioned.
Our estimate  for the universal ratio  
 between  these amplitudes: 
 $a^{+sc}_{\xi}/a^{+sc}_{\chi}=0.95(15)$ from the sc lattice series, and 
  $a^{+bcc}_{\xi}/a^{+bcc}_{\chi}=0.87(6)$ from the bcc lattice series,
 improves the 
 accuracy of our previous results\cite{bcsd} obtained from the analysis of
 shorter series. These results have to be compared with the FD  result 
 $a^+_{\xi}/a^+_{\chi}=0.65(5)$ obtained in Ref.\cite{babe} 
and with the HT result 
$a^{+bcc}_{\xi}/a^{+bcc}_{\chi}\approx 0.85$ of Ref.\cite{nr90,george}.

\section{ Conclusions} 

 We have extended through order $\beta^{23}$
 the HT expansions  
of the susceptibility and of the second correlation 
moment for the  spin-1/2 Ising model, on the sc and the bcc lattices. 
As a first application of 
our calculation, we have updated the direct HT estimates of 
 universal critical 
parameters of the Ising model  with
 some improvement over previous  analyses in the accuracy 
 and in the agreement 
 with the latest calculations by  approximate RG methods and by various
 numerical methods.
 
\acknowledgments 
This work has been partially supported by MURST. 
We thank Prof. A. Guttmann for his interest in our work and for making
 available to us Ref.\cite{guttp} before publication. We also thank
 Profs. L.N.Shchur, D. Stauffer and E. Vicari for their kind comments on the 
first version of this paper.
\begin{table}
\squeezetable
\caption{ The sequences of approximants  for $\beta_c$ and $\gamma$ defined 
by eq.(\ref{betazinn}) and  eq.(\ref{gammazinn}), 
respectively, and the sequences of the appropriate 
 extrapolations using alternate pairs, 
as obtained from  $\chi$ on the bcc lattice.
 For the extrapolations we have assumed that $\theta=0.504$. }
\label{tabella1}
\begin{tabular}{ccccc} 
n  &$(\beta_c)_n$ from  Eq.(\ref{betazinn})  & Extrapol. of $(\beta_c)_n $ 
&$\gamma_n$ from Eq.(\ref{gammazinn}) & Extrapol. of $\gamma_n$  \\
\hline  
18&0.1573815 & &1.244335 & \\
19&0.1573806   & &1.244174& \\
20&0.1573807   &0.1573761 & 1.244049 &1.238519\\
21&0.1573800 &0.1573759   &1.243889& 1.238114 \\
22&0.1573799 &0.1573743   & 1.243760 & 1.237595 \\
23&0.1573793  &0.1573746  &1.243620& 1.237599 \\
24&0.1573791    &    0.1573739 &1.243501& 1.237475 \\
25& 0.1573787  &0.1573740 &1.243374 &1.237421\\
\end{tabular} 
\end{table}

\begin{table}
\squeezetable
\caption{ A comparison among recent estimates of the critical exponents  
 $\gamma$ and $\nu$. }
\label{tabella2}
\begin{tabular}{ccccccccc} 
&This work&Series\cite{nr90}&Series \cite{campo} 
& MC\cite{blh} & MC\cite{has}
& MC\cite{balle}& FD-exp.\cite{guida}& $\epsilon$-exp.\cite{guida}  \\
\hline  
$\gamma$& 1.2375(6)&1.237(2)&1.2371(4)&1.2372(17)&1.2367(20)&1.2353(25)
&1.2396(13)&1.2380(50)\\
$\nu$&0.6302(4)&0.6300(15)&0.63002(23)&0.6303(6)&0.6296(7)&0.6294(10)&
 0.6304(13)&0.6305(25)\\
\end{tabular} 
\end{table}

\begin{table}
\squeezetable
\caption{ A comparison among recent estimates of the susceptibility universal  
amplitude  ratio $C_+/C_-$. }
\label{tabella3}
\begin{tabular}{cccccccc} 
This work&Series\cite{zinnfish,fietal}&  Eq.State\cite{campo}     
& MC\cite{ruge} & MC\cite{case}
& MC\cite{enge}& FD-exp.\cite{guida}& $\epsilon$-exp.\cite{guida}  \\
\hline  
4.762(8)&4.95(15)&4.77(2)&5.18(35)&4.75(3)&4.72(11)&4.79(10)&4.73(16)\\
\end{tabular} 
\end{table}
\begin{table}
\squeezetable
\caption{ A comparison among recent  estimates of the correlation-length 
universal amplitude  ratio $f_+/f_-$. }
\label{tabella4}
\begin{tabular}{ccccccc} 
This work&Series\cite{zinnfish,fietal}&Eq.State\cite{campo} 
 & MC\cite{ruge} & MC\cite{case}
&  FD-exp.\cite{heit}& $\epsilon$-exp.\cite{zinn}  \\
\hline  
1.963(8)&1.96(1)&1.961(7)&2.06(1)&1.95(2)&2.013(28)&1.91\\
\end{tabular} 
\end{table}

\end{document}